\documentclass[prl,twocolumn,graphics]{revtex4}
\usepackage{graphicx}
\input{epsf.sty}

\begin{document}

\title{Texture in the Superconducting Order Parameter of CeCoIn$_5$ Revealed by Nuclear Magnetic Resonance}

\author{K.~Kakuyanagi$^1$, M.~Saito$^{1}$, K.~Kumagai$^1$, S.~Takashina$^2$, M.~Nohara$^2$, H.~Takagi$^2$, and Y.~Matsuda$^3$}
\affiliation{$^1$Division of Physics, Graduate School of Science, Hokkaido University,  Sapporo 060-0810, Japan }%

\affiliation{$^2$Department of Advanced Materials Science, University of Tokyo, Kashiwanoha, Kashiwa, Chiba 277-8581, Japan}%
\affiliation{$^3$Institute for Solid State Physics, University of Tokyo, Kashiwanoha, Kashiwa, Chiba 277-8581, Japan}%


\begin{abstract}

We present a $^{115}$In NMR study of the quasi two-dimensional heavy-fermion superconductor CeCoIn$_5$ believed to host a Fulde-Ferrel-Larkin-Ovchinnkov (FFLO) state.   In the vicinity of the upper critical field and with a magnetic field applied parallel to the $ab$ plane, the NMR spectrum exhibits a dramatic change below $T^*(H)$ which well coincides with the position of reported anomalies in specific heat and ultrasound velocity.  We argue that our results provide the first microscopic evidence for the occurrence of a spatially modulated superconducting order parameter expected in a FFLO state.   The NMR spectrum also implies an anomalous electronic structure of vortex cores.

\end{abstract}

\pacs{}

\maketitle

	A myriad of fascinating properties have been proposed for unconventional superconductors in the presence of a strong magnetic field.    Among the possible exotic superconducting (SC) phases, a spatially nonuniform SC state originating from the paramagnetism of conduction electrons has become a subject of intense theoretical investigation after the pioneering work by Fulde and Ferrel and Larkin and Ovchinnikov (FFLO) in the mid-1960's \cite{fflo}.   In spin singlet superconductors, the destruction of superconductivity by a magnetic field can be achieved in two distinct ways. Cooper pairs may break up either because the spin of the conduction electron is coupled to the magnetic field (Pauli paramagnetism) or because the latter effects also the electronic orbital angular momentum (vortices).   A novel SC phase was predicted by FFLO when Pauli pair-breaking dominates over the orbital effect\cite{reiner,shimahara,gg,tachiki,budzin,agterberg,adachi,maki}.  In the FFLO state, pair-breaking due to the Pauli effect is reduced by the formation of a new pairing state ({\boldmath$k$}$\uparrow$,  {\boldmath $-k+q$}$\downarrow$) with $\mid${\boldmath $q$}$\mid$  $\sim 2 \mu_B H / \hbar v_F $ ($v_F$ is the Fermi velocity) between exchange-split parts of the Fermi surface, instead of ({\boldmath $k$}$\uparrow$,  {\boldmath $-k$}$\downarrow$)-pairing in ordinary superconductors.  In other words,  spin up and spin down electrons can only stay bound if the Cooper pairs have a drift velocity in the direction of magnetic field.   As a result, a new SC state with spatially oscillating order parameter and spin polarization with a wave length of the order of $2 \pi/|{\bf q}|$, which is comparable to the coherence length $\xi$, should appear in the vicinity of upper critical field $H_{c2}$.   

	The issue of the actual observation of the FFLO phase has only been addressed more recently especially in the last several years.   Although several type-II superconductors (including heavy fermion and organic compounds) have been proposed as likely candidates for the observation of the FFLO state,  subsequent research has called the interpretation of the data into question  \cite{gloos}.  No solid evidence, which is universally accepted as proof of the FFLO state, has turned up.   In this context, the case of CeCoIn$_5$ has aroused great interest, because several measurements have led to a renewed discussion of a possible high field FFLO state \cite{radovan,bianch,watanabe,capan}.   CeCoIn$_5$ is a new type of heavy fermion superconductor with quasi 2D electronic structure \cite{pet}  and is identified as an unconventional superconductor with, most likely, $d$-wave gap symmetry  \cite{mov,izawa,kohori,curro,pene}.   Very recent heat capacity measurements revealed that a second order phase transition takes place at $T^*(H)$ within the SC state in the vicinity of the upper critical field with {\boldmath $H$} parallel to the $ab$-plane $H_{c2}^{\parallel}$ at low temperatures  \cite{radovan,bianch}.   The transition line branches from $H_{c2}^{\parallel}$-line and decreases with decreasing $T$, indicating the presence of a novel SC phase (Hereafter we refer to the phase below $T^*(H)$ as the {\it high field SC phase}).   In the inset of Fig.~1 the  $H-T$ phase diagram for CeCoIn$_5$ is illustrated in the vicinity of $H_{c2}^{\parallel}$ and at low temperatures.     Subsequent ultrasound investigation  revealed the collapse of the flux line lattice tilt modulus \cite{watanabe}, and the thermal conductivity  reported a pronounced anisotropy \cite{capan}  in the high field SC phase below $T^*(H)$.   Both measurements were presented in support of the FFLO nature.  Thus, as new results accumulate, there is a growing experimental evidence that the FFLO state may indeed be realized in the high field SC phase of CeCoIn$_5$.   

	CeCoIn$_5$ appears to meet in an ideal way the strict requirements placed on the existence of the FFLO state.  First, an extremely high $H_{c2}^{\parallel}$ ($\sim$ 12~T at $T$=0)  is favorable for the occurence of the FFLO state because then the Pauli effect may overcome the orbital effect.   Pauli-limited superconductivity is in fact supported by the fact that the phase transition from SC to normal metal at the upper critical fields is of first order below $\sim$1.3~K  \cite{izawa,1st}.  Second, it is in the extremely clean regime.  Third,  $d$-wave pairing symmetry greatly extends the stability of the FFLO state with respect to a conventional superconductor \cite{maki}.  

	While these experimental and theoretical results make the FFLO scenario a very appealing one for CeCoIn$_5$, there is no direct experimental evidence so far which verifies the spatially nonuniform SC state expected in FFLO state.  A central matter related to this issue is the quasiparticle structure in the high field SC phase.  Therefore a powerful probe of the quasiparticle excitations in the high field SC phase is strongly required to shed light on this subject.  NMR is particularly suitable for the above purpose because NMR can monitor the low energy quasiparticle excitations sensitively.   Here we present the NMR spectrum in the vicinity of $H_{c2}^{\parallel}$ to extract  microscopic information on the quasiparticle structure for the first time.  The spectrum we observed in the high field SC phase is quite unique, and we will argue that our results provide the first microscopic evidence for the occurrence of a spatially inhomogeneous SC state expected in a FFLO state.
	

\begin{figure}[t]
\begin{center}
	\includegraphics[scale=0.6]{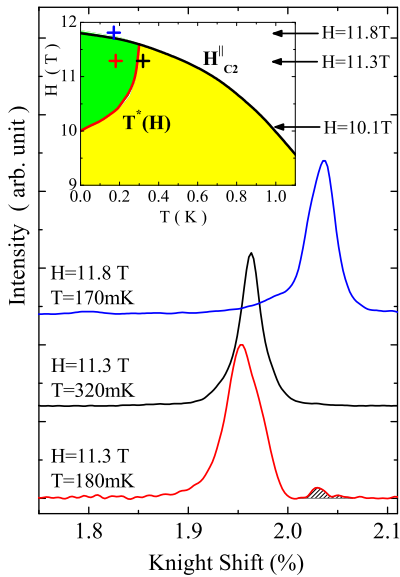}\\
	\caption{ Inset: Experimental $H-T$ phase diagram for CeCoIn$_5$ below 1~K  in {\boldmath $H$}$\parallel a$-axis.  The second order transition line $T^*(H)$  (red solid line) and   $H_{c2}^{\parallel}$  (black solid line) are obtained from Ref. \cite{capan}.  The transition at $H_{c2}^{\parallel}$ is of first order in this $T$-range.  The region shown by green depicts the high field SC phase discussed in the text.   Horizontal arrows indicate the magnetic fields at which the NMR spectrum was measured.    Main panel: $^{115}$In-NMR spectra outside, slightly above $H_{c2}^{\parallel}$ (blue  line),  slightly above $T^{*}$ (black), and well inside (red)  the high field SC phase.   The resonance feature at higher frequency is marked by hatch.  $(H,T)$-points at which each NMR spectrum was measured are shown by  crosses in the inset.  } 

\end{center}
\end{figure}	

	 $^{115}$In ($I$=9/2) NMR measurements were performed on high quality single crystals of CeCoIn$_5$ by using a phase-coherent pulsed NMR spectrometer.    Experiments were always carried out in the magnetic field {\boldmath $H$}  parallel to the [100]-direction under the field-cooled condition.  The tetragonal crystal structure of CeCoIn$_5$ consists of alternating layers of CeIn$_3$ and CoIn$_2$ and so has two inequivalent In sites per unit cell  \cite{pet}.    We report  NMR results at the In(1) site with axial symmetry in the CeIn$_3$ layer, which locates in the center of the square lattice of Ce atoms.   The Knight shift $^{115}K$ was obtained from the central $^{115}$In-line ($\pm$1/2$\leftrightarrow \mp$1/2  transition)  using a gyromagnetic ratio of $^{115} \gamma$=9.3295~MHz/T and  by taking into account the electric quadrupole interaction.   

	Figure 1 depicts the NMR spectra outside and well inside the high field SC phase.    At $T$ slightly above $T^*(H=11.3$~T)$\simeq$300~mK,  the NMR spectrum is almost symmteric as shown by the black solid line.  Generally, the spatial distribution of the magnetic field arising from the flux line lattice structure gives rise to an asymmetric NMR spectrum \cite{kakuyanagi1}, indicating that the influence of the field distribution is negligible in the present high field region.  A most remarkable feature in the NMR spectrum well inside the high field SC phase shown by the red solid line is an appearance of a new resonance peak with small but finite intensity at higher frequency , as seen clearly at $^{115}K\simeq2.03$\%.  The intensity of the higher resonance line is about 3-5 percent of the total intensity and is nearly $T$-independent below 180~mK.  This higher resonance line is an important clue to elucidate the nature of the high field SC phase.  We stress that the occurrence of the magnetic ordering is a highly unlike source for the higher resonance in view of the large difference in the intensity of the two lines.  Should antiferromagnetic order set in,  the alternating hyperfine fields would produce two unquivalent $^{115}$In(1) sites, which gives rise to the two resonance lines with equal intensities.  

        In Fig.~1,  the NMR spectrum at $H$ slightly above $H_{c2}^{\parallel}$ is also shown by the blue line.   A noteworthy feature in the spectrum inside the high field SC phase (red solid line) is that the position of the higher resonance line within the high field SC phase  coincides well with that of the resonance line above $H_{c2}^{\parallel}$ (blue),  while the position of the lower resonance line locates close to that of the SC state above $T^*(H)$ (black).   Therefore, it is natural to deduce that the  higher resonance line originates from a normal quasiparticle regime, which is newly formed below $T^*(H)$, while the lower resonance line corresponds to the SC regime, which appears to have a similar quasipaticle structure above $T^*(H)$.    These results lead us to conclude that an appearance of the new resonance line at a higher frequency is a manifestation of a novel normal quasiparticle structure in the high field SC phase. 

     Figure 2 displays the temperature evolution of the spectra at $H$=11.3~T.    The higher resonance line grows rapidly with $T$ just below $T^{*}(H)$.   A  double peak structure with same intensity shows up at $T=$240~mK,  followed by a shoulder structure at $T=260$~mK, indicating that the intensity of the higher resonance line dominates.   Two lines merge into a single line above $T^*(H)$$\sim$300mK.   The $T$-dependences of the $^{115}K$ evaluated from the peak position is plotted in Fig.~3 \cite{memo}.   $^{115}K$ at $H$=11.3~T exhibits quite ususal $T$-dependence.  As the temperature is lowered below $T^*(H)$,  $^{115}K$ of the higher resonance line increases rapidly and coincides with $^{115}K$ above $H_{c2}^{\parallel}$ below 180~mK.  On the other hand,  below $T^*(H)$, $^{115}K$ of the lower resonance line changes slightly.   

\begin{figure}[t]
\begin{center}
	\includegraphics[scale=0.6]{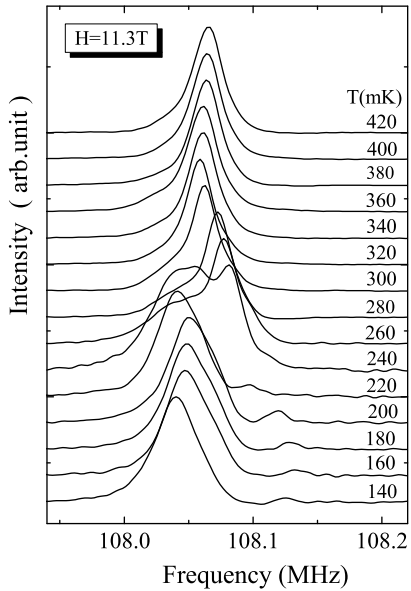}
	\caption{$^{115}$In-NMR spectra (central line of ($\pm$1/2$\leftrightarrow \mp$1/2  transition) as a function of frequency for various temperatures at  $H$=11.3~T. }
\end{center}
\end{figure}

  	So far, we have established that the high field SC phase is characterized by the formation of normal regions.  This brings us to the next question on whether the NMR spectrum below $T^{*}(H)$ is an indication of a FFLO phase.   It has been predicted that in a FFLO phase the SC order parameter exhibits one-dimensional spatial modulations along the magnetic field,  forming planar nodes that are periodically aligned perpendicularly to the flux lines.   Therefore, the formation of the normal regions is consistent with a phase expected in a FFLO state.   We will  show that the NMR spectra just below $T^*(H)$ in Fig.~2 can be accounted for by considering such planar structures.  

         The field induced layered structure expected in a FFLO phase resembles the SC states of stacks of superconductor-normal-superconductor  (S-N-S) Josephson tunnel junctions.  In the NMR experiments, $rf$ magnetic field {\boldmath $H$}$_{rf}$ is applied perpendicular to the $dc$ magnetic field ({\boldmath $H$} $\parallel a$, {\boldmath $H$}$_{rf}$ $\parallel b$).  The shielding supercurrent currents flow passing across the planar nodes.   Because of the second order transition at $T^*(H)$,  the modulation length of the order parameter parallel to {\boldmath$H$} or the thickness of the SC layers, $\Lambda(=2 \pi/|{\bf q}|)$, diverges as, $\Lambda \propto (T^*-T)^{-\alpha}$ with $\alpha>0$, upon approaching $T^*(H)$.   Therefore $\Lambda $ will exceed the in-plane penetration length $\lambda$  in the vicinity of $T^*$.  In such a situation, $rf$ field penetrates into the normal sheets much deeper than into the SC sheets,  which results in a strong enhancement of the NMR intensity from the normal sheets.  At low temperature where $\Lambda$ becomes comparable to $\xi (\ll \lambda)$,  penetration of the $rf$ field into the normal sheets is as same as that into the SC sheets.  

\begin{figure}[t]
\begin{center}
	\includegraphics[scale=0.6]{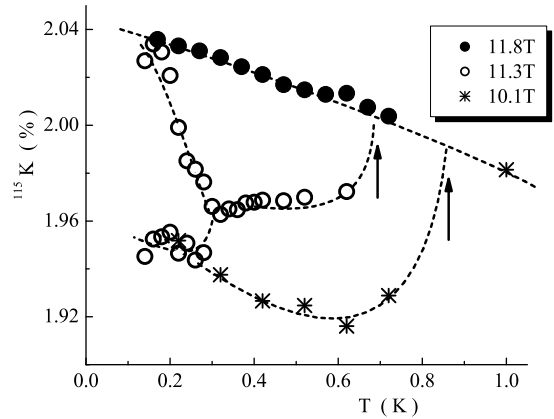}\\
	\caption{Temperature dependence of Knight shift $^{115}K$ at $H$=11.8~T ($\bullet$), 11.3~T ($\circ$) and 10.1~T ($\ast$).  The arrows show the SC transition temperatures at each corrsponding field.  The broken lines are guide for eyes.} 
\end{center}
\end{figure}

         We estimate the above effect semi-quantitatively.   Assuming a simple sinusoidal modulation of the gap function along the applied field ($x$-axis), $\Delta(x)=\Delta_0 \sin{qx}$, the spatial modulation of the $rf$ field $H_{rf}(x)$ is given as, 
	
\begin{equation}      
          H_{rf}\left(x\right) \propto {\sinh{\Lambda/2-x\over m\lambda} + \sinh{x\over \lambda} \over \sinh{\Lambda\over 2\lambda}}.
\end{equation}          
The NMR intensity is given as
	
\begin{equation}         
	I\left(k\right)=\int_0^{\Lambda} \delta\left(K\left(\Delta(x)\over T\right)-k\right) \left[H_{rf}\left(x\right)\right]^2 {\rm d}x
\end{equation}
where $K( \Delta /T)$ is the Yoshida function for the Knight shift in the SC state \cite{yoshida}.     We attempt to fit the experimental data with this formula, using  $\Lambda \over \lambda$ and $T\over \Delta_0$ as fitting parameters.   Figure 4 depicts the calculated spectrum just below $T^{*}(H)$, where we have used $T/\Delta_{0}=0.29$ and $\Lambda/\lambda=4.7$ at $T$=240~mK, and $T/\Delta_{0}=0.3$ and $\Lambda/\lambda=7.5$ at $T=$260~mK.   The spectra is also convoluted  with  a Lorentian shape for an inhomogeneous broadening.    This simple simulation, in which planar nodal structure is assumed, reproduces well  the observed spectra, and suggests that the wave length of the spatial oscillation of the SC order parameter decreases largely with lowering temperature.  Thus, the evolution of the NMR spectrum with temperature is compatible with what is expected in a FFLO phase.

\begin{figure}[t]
\begin{center}
	\includegraphics[clip,scale=0.55]{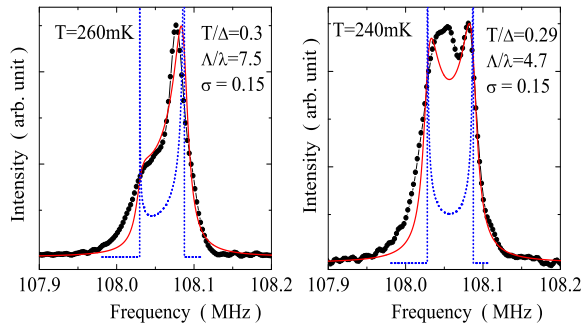}\\
	\caption{$^{115}$In-NMR spectra  ($\bullet$) at  $T$=260~mK and  $T$=240~mK.  The solid red line and blue dotted line represent the simulation spectra with and without a convolution of an inhomogeneous broadening with a Lorentian function  ($\sigma$=0.15~MHz), respectively.  For details, see the text.}
\end{center}

\end{figure}

	We finally discuss the nature of the quasiparticle structure inferred from the NMR spectrum.   The intensity of the higher resonance line indicates that only a few percent of the total volume is occupied by a newly formed normal qusiparticle region well below $T^*(H)$.  Furthermore the presence of two well-separated NMR lines implies that the quasiparticle excitation around the planar nodes is spatially localized.  We therefore speculate that the spatial dependence of the order parameter along the magnetic field may be Bloch wall-like or rectanguar, rather than sinusoidal far below $T^*(H)$.  These results call for further theoretical investigations on the real space structure of the SC order parameter.  

	  We note that a peculiar electronic structure of vortex cores in CeCoIn$_5$ is also inferred from the present results.   A double peak structure in the NMR spectra directly indicates that the Knight shift within the vortex core deviates from that in the normal quasiparticle sheets.  This implies that the vortex core is to be distinguished from the normal state above $H_{c2}^{\parallel}$, a feature in sharp contrast to conventional superconductors, where the Knight shift within the core coincides with the Knight shift in the normal state.  What is the reason behind this unusual structure of the vortex core?  Since in CeCoIn$_{5}$ $H_{c2}^{\parallel}$ is limited by Pauli paramagnetic effect, the area occupied by vortex cores can be much smaller than what is estimated from  $H/H_{c2}^{\parallel}$.  Hence, even just below $H_{c2}^{\parallel}$, the vortex cores are associated with a large spatial oscillation of the SC order parameter.  We recall that a strong reduction of the quasiparticle density of states within vortex cores has been reported in high-$T_{c}$ cuprates \cite{kakuyanagi1,pan}, and discussed in terms of the strong enhancement of the antiferromagnetic correlation within cores \cite{kakuyanagi2}.  A similar situation may be present in CeCoIn$_5$.  Interestingly, Nernst effect measurements in the latter \cite{bel} indicate that the difference of entropy between the vortex core and the superconducting environment is unusually small and are therefore compatible with a reduced density of states in the vortex core. Moreover,
Strongly enhanced antiferromagnetic correlation in  CeCoIn$_{5}$ is inferred from the $T$-shift of Knight shift in the normal state above $H_{c2}^{\parallel}$, which increases with decreasing $T$, as is evident from Fig.~3.  This behavior is notably different from that expected in the Fermi liquid model, which predicts the $T$-independent Knight shift.   This non-Fermi liquid behavior has been discussed in the light of the incipient antiferromagnetism with the quantum critical point in the vicinity of the upper critical field \cite{QPC}.  These results call for further investigations of the vortex core structure in the presnce of strong antiferromagnetic correlation.

	To conclude,  $^{115}$In NMR spectrum in CeCoIn$_{5}$ exhibits a dramatic change in the vicinity of $H_{c2}^{\parallel}$.  Below $T^*(H)$ a new resonance line appears at higher frequency, which can be attributed to the normal quasiparticle sheets formed in the SC regime.     On the basis of the NMR spectrum, we were able to establish a clear evidence of the spatially inhomogeneous SC state at high field and low temperatures, precisely as expected in a FFLO state.  The NMR spectrum also indicate that the vortex core structure of CeCoIn$_5$ appears to be  markedly different from that of ordinary superconductors.

	We thank  K.~Behnia, Y. Furukawa, R.~Ikeda, K.~Ishida, A. Kawamoto, T.~Kita, K.~Machida, K.~Maki. H.~Shimahara, M.~Takigawa, A.~Tanaka, and T.~Watanabe for stimulating discussions.


\begin{thebibliography}{99}
\bibitem{memo} We obtained the Knight shift by taking into account the shift due to the large quadrupole interaction with  the quadrupole frequency, $\nu _Q$=8.173~MHz and asymmetric paramer, $\eta$=0 for the In (1) site.
\bibitem{yoshida}  We assumed a local DOS with a $d$-wave SC gap.
\end{thebibliography}
\end{document}